\documentclass[preprint2]{aastex}
\usepackage{dcolumn}
\usepackage{bm}
\usepackage{lscape}
\usepackage[dvips]{epsfig}
\usepackage{amsmath}
\usepackage{amssymb}
\usepackage{amsthm}
\usepackage{array}
\usepackage{multirow}
\usepackage{graphicx}
\usepackage{amsfonts}
\usepackage{eucal}
\usepackage[]{appendix}

\begin{document}
\title{\large Concurrent application of ANC and THM to assess the $^{13}{\rm C}(\alpha,n)^{16}{\rm O}$ absolute cross section at astrophysical energies and possible consequences for neutron production in low-mass AGB stars}

\author{O. Trippella}
\affil{Department of Physics and Geology, University of Perugia, Perugia, Italy}
\affil{Istituto Nazionale di Fisica Nucleare, Section of Perugia, Perugia, Italy}
\email{oscar.trippella@pg.infn.it}

\author{M. La Cognata}
\affil{Istituto Nazionale di Fisica Nucleare, Laboratori Nazionali del Sud, Catania, Italy}



\begin{abstract}
The $^{13}{\rm C}(\alpha,n)^{16}{\rm O}$ reaction is considered to be the main neutron source responsible for the production of heavy nuclides (from ${\rm Sr}$ to ${\rm Bi}$) through slow $n$-capture nucleosynthesis ($s$-process) at low temperatures during the asymptotic giant branch (AGB) phase of low mass stars ($\lesssim 3-4\;{\rm M}_{\odot}$, or LMSs). In recent years, several direct and indirect measurements have been carried out to determine the cross section at the energies of astrophysical interest (around $190\pm40\;{\rm keV}$). However, they yield inconsistent results causing a highly uncertain reaction rate and affecting the neutron release in LMSs. In this work we have combined two indirect approaches, the asymptotic normalization coefficient (or ANC) and the Trojan Horse Method (THM), to unambiguously determine the absolute value of the $^{13}{\rm C}(\alpha,n)^{16}{\rm O}$ astrophysical factor. Therefore, we have determined a very accurate reaction rate to be introduced into astrophysical models of $s$-process nucleosynthesis in LMSs. Calculations using such recommended rate have shown limited variations in the production of those neutron-rich nuclei (with $86\leq A\leq 209$) receiving contribution only by slow neutron captures.
\end{abstract}

\section{Introduction \label{sec-intro}}
Slow neutron captures, making up the so-called $s$-process, are responsible for the production of about $50\%$ of nuclei heavier than iron. They take place during the ${\rm He}$- and ${\rm C}$-burning phases of massive stars for the production of nuclides between iron and strontium \citep[$60 \lesssim A \lesssim 90$, ][]{PIG10} and in the ${\rm He}$-burning layers of low- and intermediate-mass asymptotic giant branch (AGB) stars for the main component \citep[between ${\rm Sr}$ and ${\rm Bi}$, ][]{GAL98}. In this paper we focus our attention on this latter astrophysical site, specifically the thermally pulsing AGB phase of low-mass stars \citep[LMSs, ${\rm M}_{\bigstar} \lesssim 3-4 \;{\rm M}_{\odot}$\footnote{We consider as the limit between low- and intermediate-mass stars, the mass of those stars that do not have the temperature high enough to activate the ${\rm CNO}$ cycle at the base of the convective envelope (hot bottom burning) and the  $^{22}{\rm Ne}(\alpha,{\rm n})^{25}{\rm Mg}$ (see later discussion in this section) neutron source is only partially activated always because of low temperature reached.}, ][]{BUS01,STR03}. During these stages, a star is characterized by a structure made of a ${\rm C}$-${\rm O}$ degenerate core surrounded by two shells, the inner composed by helium and the outer hydrogen rich, burning alternatively \citep{IBE83}. As shell ${\rm H}$ burning proceeds while the ${\rm He}$ shell is inactive, the mass of the ${\rm He}$ increases and attains higher densities and temperatures. As a consequence, ${\rm He}$-burning in the shell is 
temporarily activated by thermonuclear runaway flash events generating convective instabilities (or thermal pulses) thanks to sudden temperature enhancements. These stars undergo repeated mixing episodes (the so-called third dredge-up, TDU) of material below the H-burning shell, where ${\rm He}$-burning and slow neutron captures occur, bringing the fresh material just produced toward the stellar surface \citep{BUS99,HER05,KAP11}. In this astrophysical scenario, the main neutron source has been identified in the $^{13}{\rm C}(\alpha,n)^{16}{\rm O}$ reaction activated in radiative conditions during the quiet phases between two subsequent thermal instabilities at a temperature of about $0.9 \times 10^8\;{\rm K}$ \citep{GAL88}. A second neutron exposure is due to $^{22}{\rm Ne}(\alpha,{\rm n})^{25}{\rm Mg}$ reaction during the convective instabilities of helium shell providing more intense neutron fluxes. However, this is only marginally activated because of typical low temperature, only about $2.5 \times 10^8\;{\rm K}$, of stars less massive than $3\;{\rm M}_{\odot}$ \citep{STRA95}.

The typical neutron densities for $s$-process in LMSs provided by the $^{13}{\rm C}(\alpha,n)^{16}{\rm O}$ reaction are about $10^6-10^8\;{\rm n/cm}^3$. Existing direct measurements, collected in the European Compilation of Reactions Rates for Nuclear Astrophysics (NACRE) by \citet{ANG99} and the subsequent updated version by \citet{XU13} (hereafter NACRE II), stop at the minimum value of about $280\;{\rm keV}$ \citep{DRO93}, whereas the region of astrophysical interest, the so-called Gamow window \citep{ROL88,ILI07}, corresponds to about $150-230\;{\rm keV}$ at a temperature of $10^8\;{\rm K}$. At low temperature, the main uncertainty source is represented by the presence of a resonance near the $\alpha$-threshold corresponding to the $1/2^{+}$ excited state of $^{17}{\rm O}$. The most recent works \citep{HEI08,LAC12,GUO12,LAC13,XU13,AVI15} present in the literature are oriented towards a substantial lowering of the reaction rate with respect to the one suggested by NACRE \citep{HAL97}, because it is believed that the role of the resonance mentioned above was overestimated in the past \citep{HEI08}. In this scenario, the $^{13}{\rm C}(\alpha,n)^{16}{\rm O}$ reaction plays a crucial rule determining the neutron production and the time scale of $^{13}{\rm C}$ burning at a given temperature, and influencing the possibility of a complete exhaustion of the fuel during the radiative phase \citep{CRI11}. For these reasons, its efficiency is still matter of debate, aiming at reducing the uncertainty that can reach about $300\%$ in the most interesting astrophysical region \citep{ANG99,JOH06}. 

Moreover, the $^{13}{\rm C}(\alpha,n)^{16}{\rm O}$ reaction is also activated in other different astrophysical sites \citep{JOR89}, such as at the ${\rm C}$-burning in massive stars; at the beginning of ${\rm He}$-burning phase; and when there exist a proton injection in ${\rm He}$-rich layers (e.g. central ${\rm He}$ flash in star less massive than $2\;{\rm M}_{\odot}$; for a late convective instability in nuclei of planetary nebulae at the beginning of Wolf-Rayet, type ${\rm N}$, phase showing little hydrogen and nitrogen enhancement at the stellar surface, on certain massive mass-losing stars; and accretion of ${\rm H}$-rich material on white dwarf in a binary system). The analysis of consequences of the $^{13}{\rm C}(\alpha,n)^{16}{\rm O}$ reaction in these conditions is beyond the primary purpose of this article, but it is highly desirable.

After a brief presentation of the status of the art of the measurements for the $^{13}{\rm C}(\alpha,n)^{16}{\rm O}$ reaction cross section (Sec. \ref{sec-status}), we present in Sec. \ref{sec-rea} a new approach starting from indirect experimental data to constraint the absolute normalization factor of direct measurements at high energies. The astrophysical factor is calculated and compared (Sec. \ref{sec-astro-fac}) with the most recent works present in the literature, also focusing on the electron screening effect (Sec. \ref{sec-ele-scree}). In Sec. \ref{sec-rate} we determine the recommended reaction rate for the $^{13}{\rm C}(\alpha,n)^{16}{\rm O}$ reaction to be used in the astrophysical scenarios outlined above. In particular, here we will evaluate possible consequences on the $s$-process nucleosynthesis.

\section{Status of the art \label{sec-status}}
Due to the relevance of the $^{13}{\rm C}(\alpha,n)^{16}{\rm O}$ reaction as the main neutron source for the $s$-process in LMSs, during last decades several measurements of its cross section have been performed covering different energy ranges. 
One of the most commonly adopted rate was that presented in \citet{ANG99}, which takes into account experimental cross sections determined in previous works by \citet{SEK67}, \citet{DAV68,BAI73,DRO93,BRU93}. 
Later, an unprecedented accuracy of $4\%$ was reached by \citet{HAR05}, triggered by the need to reliably subtract the background in the observation of geo-neutrinos \citep[e.g. in the KamLAND detector,][]{ARA05}. One of the most recent work on the $^{13}{\rm C}(\alpha,n)^{16}{\rm O}$ reaction by \citet{HEI08} combines a high accuracy cross section measurements down to about $300$ ${\rm keV}$ with an extensive multi-channel $R$-matrix fitting of all cross section data for the channels feeding the $^{17}{\rm O}$ states contributing to the $^{13}{\rm C}(\alpha,n)^{16}{\rm O}$ reaction in the energy range below about $500\;{\rm keV}$. 

However, many experimental difficulties exist (e.g. Coulomb suppression, electron screening effect, and a different absolute value) when performing a direct measurement at the energies relevant for astrophysics. In this context, two characteristics are evident for direct data concerning the astrophysical factor of the $^{13}{\rm C}(\alpha,n)^{16}{\rm O}$ reaction: (i) the very high uncertainties due to the prohibitively small reaction cross section in the low-energy region; (ii) data by \citet{DAV68,DRO93,HEI08} appear to be consistent with each others, but they are different from ones of \citet{BAI73} specially, for an issue connected to resolution, concerning the height of the resonances above $500\;{\rm keV}$. At the same time measurements of \citet{KEL89} and \citet{HAR05} are in agreement, but they are considerably lower, in absolute value, than those mentioned above.
 
Moreover, since direct measurements stop right at the edge of the Gamow window, several experiments using indirect methods (determining the spectroscopic factor and/or the asymptotic normalization coefficient, ANC, for the $1/2^+$ level of $^{17}{\rm O}$ near threshold) have been performed to determine the cross section of this neutron source in the relevant energy region for astrophysics.
In \citet{KUB03}, the measurement of the ${}^{13}{\rm C}(^{6}{\rm Li},d)^{17}{\rm O}$ transfer reaction suggested a very small spectroscopic factor $S_{\alpha}= 0.01$, then re-analysed by \citet{KEE03} indicating a considerably stronger contribution, about a factor of $40$ larger, depending on the theoretical approach. The first determination of the ANC for the $^{13}{\rm C}(\alpha,n)^{16}{\rm O}$ reaction was performed by \citet{JOH06} using the ${}^6{\rm Li}({}^{13}{\rm C},{\rm d}){}^{17}{\rm O}$ sub Coulomb $\alpha$-transfer reaction. The authors obtained a value of the squared Coulomb-modified ANC $\left( \tilde{C}_{\alpha{}^{13}{\rm C}}^{{}^{17}{\rm O}(1/2^+)}\right)^2=0.89\pm 0.23$ ${\rm fm}^{-1}$. This result, appearing to be inconsistent with the emerging scenario of the $^{13}{\rm C}(\alpha,n)^{16}{\rm O}$ reaction, was recently revisited in the paper \citet{AVI15}. In both cases, the $\alpha$-transfer ${}^{6}{\rm Li}({}^{13}{\rm C},d){}^{17}{\rm O}$ reaction was performed with a $8\;{\rm MeV}$ $^{13}{\rm C}$ beam in inverse kinematics to achieve the lowest possible energy in the center-of-mass system. Taking into account target deterioration, the squared Coulomb-modified ANC becomes $3.6 \pm 0.7\;{\rm fm}^{-1}$, a factor of four larger than the one suggested in the first analysis of the same measurement.

\citet{PEL08} and \citet{GUO12} also determined both spectroscopic factor, $S = 0.29 \pm 0.11$ and $0.37 \pm 0.12$; and the ANC $\left(\tilde{C}_{\alpha{}^{13}{\rm C}}^{{}^{17}{\rm O}(1/2^+)}\right)^2= 4.5 \pm 2.2\;{\rm fm}^{-1}$ and $4.0 \pm 1.1 \;{\rm fm}^{-1}$, respectively. Therefore, independent ANC experiments, using different transfer reactions and theoretical approaches, seems to indicate values for $C^2$ in the range $0.89-4.5\;{\rm fm}^{-1}$ (see Tab. \ref{tab-gam}); where the very low value suggested by \citet{JOH06}, as already explained above, was recently revised in \citet{AVI15}.



\begin{table*}
\begin{center}
\caption{\label{tab-gam}Summary of widths and ANC values for the $1/2^{+}$ state of $^{17}{\rm O}$ close to the $^{13}{\rm C}-\alpha$ threshold reported in the literature.}

\begin{tabular}{lccc}
\\
\hline\hline
Reference 		& $\Gamma_n\;{\rm(keV)}$& $ $ & ANC $({\rm fm}^{-1})$\\
\hline
\citet{FOW73}    & $124$& $  $ &  \\
\citet{TIL93}		& $124 \pm 12$& $  $ &  \\
\citet{SAY00}		& $162.37$& $  $ &  \\
\citet{JOH06}    & $124\pm12$& $  $ &  $0.89\pm0.23$ \\
\citet{PEL08}    & $124\pm8$& $  $ &  $4.5\pm2.2$ \\
\citet{HEI08}		& $158.1$& $  $ &  \\
\citet{LAC12}		& $83^{+9}_{-12}$& $  $ & $6.7^{+0.9}_{-0.6}$  \\
\citet{GUO12}    & $124$& $  $ &  $4.0\pm1.1$ \\
\citet{LAC13}		& $107 \pm 5_{\rm stat}{}^{+9}_{-5}{}_{\rm norm}$& $  $ &  $7.7 \pm 0.3_{\rm stat}{}^{+1.6}_{-1.5}{}_{\rm norm}$\\
\citet{FAE15}\tablenotemark{a} \tablenotetext{a}{These values are also used in this paper (see Sec. \ref{sec-rea}).}		& $136 \pm 5 $ & $  $ &  \\
\citet{AVI15}   & $  $ &  & $3.6\pm0.7$ \\
\hline
\end{tabular}
\end{center}
\end{table*}

The experiment performed applying the indirect Trojan Horse Method \citep[hereafter THM,][]{SPI11} at the Tandem-LINAC facility of the Florida State University deserves a separate discussion, because it extends over a broad energy range between $-0.3$ and $1.2\;{\rm MeV}$, therefore both negative $E_{c.m.}$ and the region covered by direct measurements are explored in a single measurement. 
The $^{13}{\rm C}(^{6}{\rm Li},n^{16}{\rm O})d$ reaction was studied in quasi-free kinematic conditions (the deuteron inside the $^{6}$Li beam is considered as a spectator to the three-body reaction) in order to deduce the astrophysical $S(E)$ factor of the $^{13}{\rm C}(\alpha,n)^{16}{\rm O}$ reaction free of Coulomb suppression and electron screening effects. 
Two peaks at positive $E_{c.m.}$ were used for normalization to direct measurements, while the ANC was extracted for the resonance below the $\alpha$ emission threshold. The normalization of indirect data to direct measurements represents the most crucial and sensible phase of THM off-line analysis using an energy region covered by both types of experiments. 
This procedure strongly depends by uncertainties affecting direct measurements and often represents the higher source of error for the THM astrophysical factor. The observable partial width $\Gamma_n^{1/2^+}$ of the $-3$ ${\rm keV}$ resonance was obtained from the $R$-matrix fit of the same $S$-factor, yielding to a value of $107 \pm 5_{\rm stat}{}^{+9}_{-5}{}_{\rm norm}$ ${\rm keV}$, larger than the one obtained in our preliminary analysis \citep{LAC12} $\Gamma_n^{1/2^+}=83^{+9}_{-12}\;{\rm keV}$. The new result is slightly smaller than the value usually adopted in the literature, $124\pm12$ ${\rm keV}$ \citep{TIL93}, and of the one reported in \citet{HEI08}, $158\;{\rm keV}$. As mentioned before, the THM approach allowed us also to extract the Coulomb-modified ANC $\tilde{C}_{\alpha{}^{13}{\rm C}}^{{}^{17}{\rm O}(1/2^+)}$ of the $-3\;{\rm keV}$ resonance from the same data sets (at odds with other direct and indirect measurements), from the HOES (half-of-energy-shell) $R$-matrix fitting of the THM data. Then, it was the first time that THM was used to extract the ANC of a sub threshold resonance. In detail, we obtained the $\left( \tilde{C}_{\alpha{}^{13}{\rm C}}^{{}^{17}{\rm O}(1/2^+)}\right)^2=7.7 \pm 0.3_{\rm stat}{}^{+1.6}_{-1.5}{}_{\rm norm}\;{\rm fm}^{-1}$, as just diffusely discussed in \citet{MUK99} and \citet{LAC12}. The resulting ANC, obtained with the complete THM data set, is in agreement with our preliminary value $6.7^{+0.9}_{-0.6}\;{\rm fm}^{-1}$, within the uncertainties.

Moreover, a new calculation of the $^{13}{\rm C}(\alpha,n)^{16}{\rm O}$ reaction rate is available in the updated version of NACRE compilation \citep[NACRE II, ][]{XU13}. It considers, in addition to the data already mentioned in \citet{ANG99}, the direct measurements of \citet{HAR05} and \citet{HEI08} to extend the energy range up to $E_{c.m.} \sim 6\;{\rm MeV}$ and indirect ones \citep{KUB03,KEE03,PEL08,GUO12,LAC12} to take into account the contribution of the state near the $\alpha$-threshold. At low temperature, the ensuing reaction rate is significantly reduced with respect to their previous determination \citep{ANG99} because of the choice of a very steep $S$-factor assumed for the $1/2^{+}$ resonance contributions. However, the corresponding uncertainties are very high, up to $+36\%$ and $-28\%$ around $0.09\;{\rm GK}$, and they are connected to the normalization procedure between NACRE II calculation and direct data. This fact clearly called for further investigations of the absolute value of the astrophysical $S$-factor.

The ANC value by \citet{AVI15}, which is the most precise calculation to date, is now compatible with results of \citet{PEL08} and \citet{GUO12}, but a discrepancy is still evident \citep[see Fig. 4 in ][and Tab. \ref{tab-gam} for a detailed comparison between several works present in literature]{AVI15} with the two measurements performed applying the THM \citep{LAC12,LAC13}, pointing at some issue that is worth to be investigated to supply a reliable reaction rate for astrophysical applications. The understanding of such discrepancy is among the reasons behind this work, aiming at proposing a consistent description of the $^{13}{\rm C}(\alpha,n)^{16}{\rm O}$ reaction. As already discussed before, the main issue affecting THM measurement is the normalization procedure to direct data at high energies.

In a recent work \citep{FAE15}, levels of $^{17}{\rm O}$ close to the $^{13}{\rm C}+\alpha$ threshold were studied by investigating the $^{19}{\rm F}({\rm d},\alpha)^{17}{\rm O}$ reaction to measure  their excitation energies and widths with high accuracy because of their astrophysical importance.
The new results recommended in \citet{FAE15} for the $1/2^{+}$ state are $E^* = 6.3634 \pm 0.0031\;{\rm MeV}$ and $\Gamma = 0.136 \pm 0.005\;{\rm MeV}$, significantly improving the uncertainties and modifying the values  assumed so far. The new excitation energy is $7.4\;{\rm keV}$ higher than the adopted value \citep{TIL93}, while the threshold value is almost unchanged and it is considered to be at $6.359\;{\rm MeV}$. As a first consequence, the $1/2^{+}$ level of $^{17}{\rm O}$ is now centred at about $4.7 \pm 3\;{\rm keV}$ above the threshold. This state can no longer be considered a sub threshold resonance, but it is now more properly called threshold resonance.

In order to have an idea of the ambiguous situation concerning the $1/2^{+}$ state, a series of neutron reduced width present in literature is listed in Tab. \ref{tab-gam}. There exists a large spread for possible $\Gamma_n$ values ranging between $83$ and $162\;{\rm keV}$ and a clear solution is far to be found. For all these reasons the contribution to the astrophysical factor of the broad resonance located near the $^{13}{\rm C}(\alpha,n)^{16}{\rm O}$ threshold still remains matter of debate and we address this issue in next sections. In this context, the THM represented an alternative and complementary approach in order to discriminate among precedent experiment performed via indirect techniques showing non compatible results for the $^{13}{\rm C}(\alpha,n)^{16}{\rm O}$ reaction. Moreover, the value of ANC, for which recently seems to exist a converging value around $4\;{\rm fm}^{-1}$, together with the THM data could represent a promising starting point to constrain the absolute value of normalization trying to single out among the several direct data present in literature.

Our goal is to overturn the present paradigm in the application of indirect approaches. We will make a synergic use of THM and ANC to assess the absolute normalization of direct data at high energies, greatly reducing systematic errors affecting the reaction rate.

\section{Reanalysis of THM data \label{sec-rea}}
The two most recent papers \citep{FAE15,AVI15} presented in the previous section described the $1/2^{+}$ excited state of $^{17}{\rm O}$ near the $^{13}{\rm C}+\alpha$ threshold with high accuracy, with crucial consequences on the current scenario outlined in \citet{LAC13} for the astrophysical factor of the $^{13}{\rm C}(\alpha,n)^{16}{\rm O}$ reaction. A reanalysis of data obtained through THM becomes, then, necessary in order to: (i) verify any variation of the ANC value as a consequence of the changed resonance energy and width of $1/2^+$ level near the $^{13}C+\alpha$ threshold; (ii) use the ANC and THM data aiming to validate direct measurements.

From a theoretical point of view, we can not extract the ANC value from THM data as just done for the first time in \citet{LAC12} and then in \citet{LAC13} because, as claimed by \citet{FAE15} the resonant state mentioned above is not a sub threshold resonance, but it is centered at $4.7\;{\rm keV}$. So, an extension of the technique is required in order to obtain the $\tilde{C}_{\alpha{}^{13}{\rm C}}^{{}^{17}{\rm O}(1/2^+)}$ in the positive energy region. As it is demonstrated in the Appendix, this is possible considering  equation \ref{anc-2} for captures to unbound states. This formula was used for the calculation of the ANC starting from the reduced width extracted from THM data \citep{LAC13} and from the resonance energy determined by \citet{FAE15}; a very large value, around $12.1^{+1.9}_{-1.6}\;{\rm fm}^{-1}$ is retrieved, significantly departing from the values in the literature: see third column of Tab. \ref{tab-gam} for a comparison with other calculations in the past. 
Because THM data require to be anchored to direct measurements at higher energies in order to obtain absolute values, the uncertainty affecting direct data in the normalization region could be the cause of the discrepancy between ANC calculated from THM \citep{LAC12,LAC13} and the ones measured by transfer reactions \citep{PEL08,GUO12,AVI15}. Since, recently, very accurate values of ANC are available above, we choose to reverse the usual normalization procedure, charactering THM experiments, adopting the $(\tilde{C})^2$ of literature scaling the THM data to the sub-threshold resonance. In this way, it will be possible to assess the correct normalization of direct data using THM.

Secondly, the neutron reduced width of \citet{FAE15} is bigger than the one shown in \citet{LAC13}, where authors obtained $\Gamma_n = 107 \pm 5_{\rm stat}{}^{+9}_{-5}{}_{\rm norm}$ assuming a different location, $E_{c.m.}=-3\;{\rm keV}$ for the $1/2^{+}$ state. In particular, experiments performed via the THM do not provide an accurate measure of the position for resonances because of typical value of energy resolution preventing us to discriminate between the two scenarios described above. On the other hand, \citet{AVI15} provided the most precise value for ANC, but it is considerably smaller than the one suggested by the precedent analysis of THM measurement.

In this context, we decide to assume the most precise value for the ANC given by \citet{AVI15} and to calculate the reduced width $\gamma_{\alpha}$ applying equation \ref{anc-2}. On the other side, in order to constrain the neutron partial width $\Gamma_n$ we use the precise evaluation of \citet{FAE15}, so both $\gamma{\rm s}$ are defined at the beginning of our approach. These values will be used in the next section with the aim to determine the normalization factor in order to establish a connection between THM indirect data and direct measurements. 

In particular, we want now to verify possible changes when we assume these new parameters in the calculation of the astrophysical factor for the $^{13}{\rm C}(\alpha,n)^{16}{\rm O}$ reaction starting from the THM measurements of \citet{LAC13} and, at the same time, to identify which direct datasets better agrees with indirect data obtained via THM.

\subsection{A New Normalization Procedure for THM Experiments \label{sec-norm}}
\begin{figure*}
\centering
\includegraphics[width=0.85\linewidth]{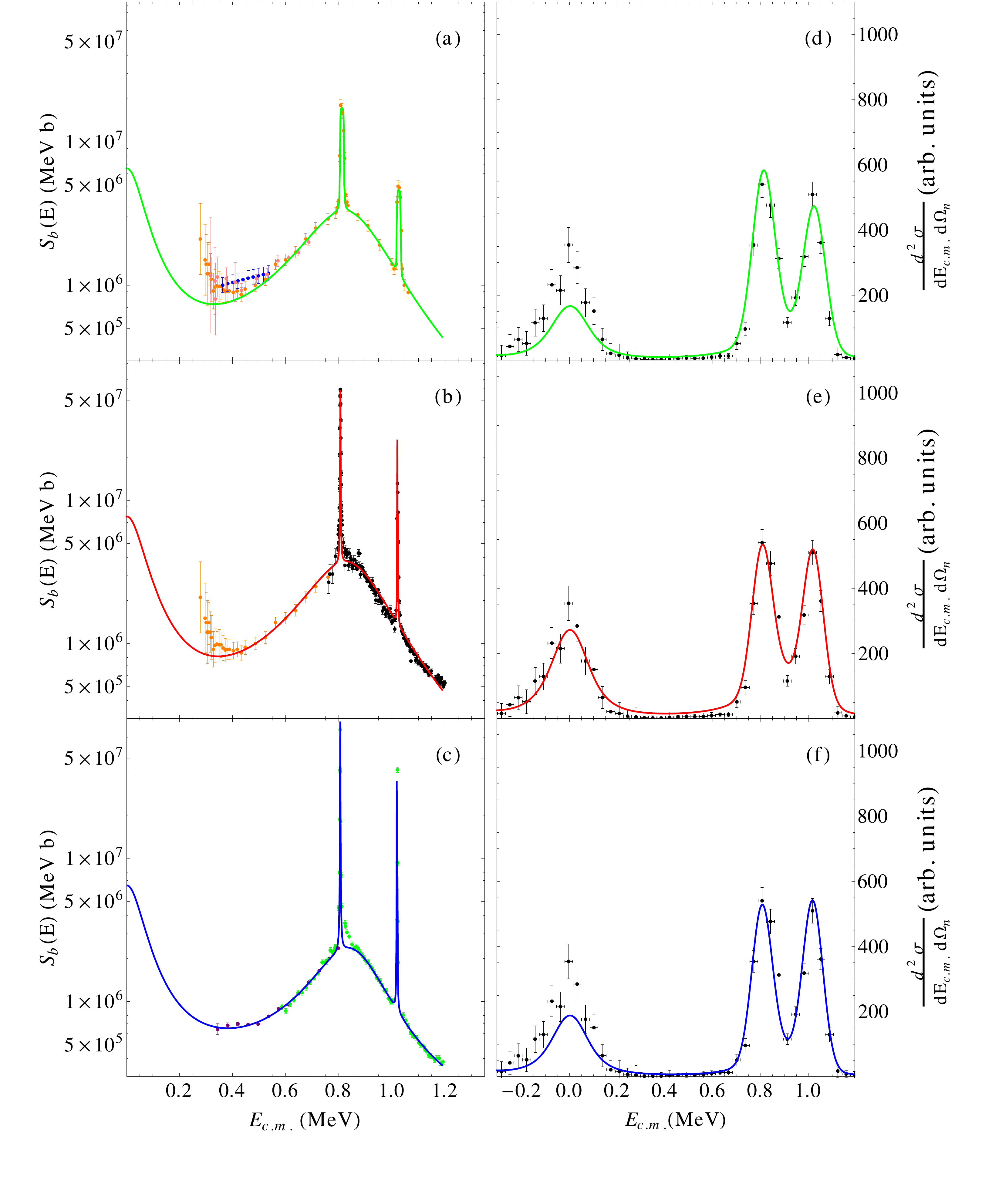}
\caption{\textit{Left panels} $-$ $R$-matrix astrophysical factor calculated assuming three different direct data sets for normalization procedure: (a) blue \citep{DAV68}, orange \citep{DRO93}, and pink \citep{HEI08} symbols; (b) black \citep{BAI73} and orange \citep{DRO93} for the low-energy region ($E_{c.m.} \lesssim 0.75\;{\rm MeV}$); (c) purple \citep{KEL89} and green \citep{HAR05}. \textit{Right panels} $-$ HOES $R$-matrix fit of the THM data presented in \citet{LAC13} (solid black symbols) adopting the same resonance parameters used in the corresponding left panel. According to $\tilde{\chi}^2$ value, the best choice for normalization of THM data corresponds to panels (b) and (e).\label{dir-ind}}
\end{figure*}

As anticipated in previous sections, we aim to use THM data to give information on the absolute value of the $S(E)$-factor for the $^{13}{\rm C}(\alpha,n)^{16}{\rm O}$ reaction, making it possible to discriminate between the several series of direct experimental data present in the literature and obtained with different techniques, degrees of accuracy and covering various energy regions. The HOES $d^{2}\sigma/dE_{c.m.}d\Omega_{d}$ cross section, obtained through the THM technique, is at present expressed in arbitrary units, making it necessary to introduce of a normalization factor representing the only free parameter to match the modified $R$-matrix calculation with indirect data, where available. For these reasons, the THM is not considered an alternative approach, but a complementary one with respect to direct measurements. As it has been shown in several works \citep{LAC11,LAC12}, normalization can be achieved by extending the indirect measurement to an energy region where directly measured resonances are available, fixing the reduced widths to match the values in the literature and determining a scaling factor between direct and THM data. However, this procedure often represents the largest source of uncertainty for the indirect $S$-factor, also because large uncertainties often affect direct astrophysical factor right at normalization  energies, for instance, in the case of the $^{13}{\rm C}(\alpha,n)^{16}{\rm O}$ reaction.

In this context, we performed a new analysis of THM data reported in \citet{LAC13} starting from the width and resonance energy reported by \citet{FAE15}, for the $1/2^{+}$ state of $^{17}{\rm O}$ located near the $^{13}{\rm C}+\alpha$ threshold, and the ANC value measured by \citet{AVI15}, to constrain the normalization to the direct measurement that best fits the THM data scaled to match such ANC. In Fig. \ref{dir-ind} we present three possibilities for normalization adopting different datasets of direct experiments from literature. Left panels show the astrophysical factor $R$-matrix fit superimposed on direct data measurements. In the first panel, (a), we have selected data from \citet{DAV68,DRO93,HEI08} (represented by blue, orange, and pink solid points, respectively) proving to be consistent with each other without the need to adopt any normalization, unlike what we did in our previous calculations \citep{LAC12,LAC13} following the idea of \citet{HEI08}. Below, panel (b) shows, by means of black circles, the large data set of \citet{BAI73} \citep[this one is more extended than in ][]{HEI08,LAC12,LAC13}, while we used data by \citet{DRO93} for the energy region $E_{c.m.}\leq 0.75\;{\rm MeV}$. Panel (c) displays the two datasets showing the lowest absolute values for the $^{13}{\rm C}(\alpha,n)^{16}{\rm O}$ astrophysical factor among those present in literature, in particular \citet{HAR05} (red full points) and \citet{KEL89} (purple ones). In this context, direct data from \citet{BRU93} are not shown because, as discussed in the same paper, they correspond to the cross section calculated by \citet{KEL89} multiplied by a factor of $1.17$, and they appear to be compatible with the other direct measurements only as a result of their large uncertainties.

\begin{figure}
\includegraphics[width=0.85\columnwidth]{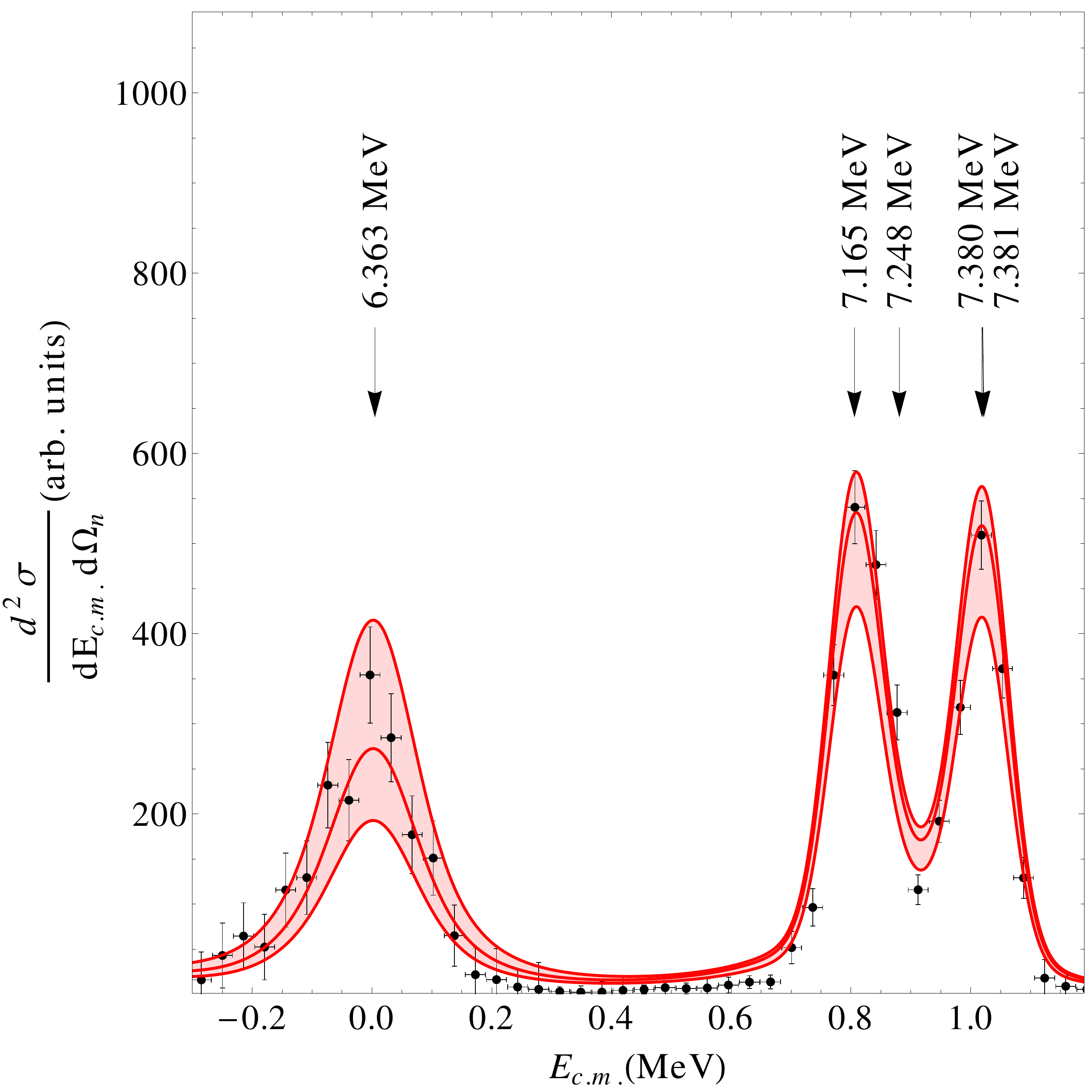}
\caption{HOES $R$-matrix fit of the THM data presented in \citet{LAC13} (solid black symbols). The middle red line is the best-fit curve and the red band represents the region allowed by statistical and normalization uncertainties. Black arrows show the centroid position of fives resonances corresponding to excited states of $^{17}{\rm O}$. Above $500\;{\rm keV}$, the parameters of the resonances were the same used in panels (b) and (e) of Fig. \ref{dir-ind}, while we adopted the partial width suggested by \citet{FAE15} for the threshold level.\label{hoes}}
\end{figure}

Right panels of Fig. \ref{dir-ind} show the comparison between modified $R$-matrix calculation and the $d^{2}\sigma/dE_{c.m.}d\Omega_{d}$ cross section determined by the THM experiment for the $^{13}{\rm C}(\alpha,n)^{16}{\rm O}$ reaction described in \citet{LAC13}. The color of curves is the same used for $R$-matrix lines in the corresponding left panel, to point out that the same parameters for all resonances were adopted for both on-energy-shell (or OES) and HOES calculations. In this phase, we do not consider upper and lower limits, but only the best fits are shown in each panel of Fig. \ref{dir-ind}. Black solid points represents the weighted sum of two datasets obtained with different $^{13}{\rm C}$ target thicknesses: specifically, $53$ and $107\;{\rm\,\mu g\,cm^{-2}}$. These two datasets were demonstrated to be consistent with each other in \citet{LAC13}, having a theoretical energy resolution (standard deviations) of $45$ and $48\;{\rm keV}$, respectively. The $d^{2}\sigma/dE_{c.m.}d\Omega_{d}$ clearly shows the presence of several resonances in the $^{13}{\rm C}-\alpha$ relative energy spectrum and more details are available in the description of Fig. \ref{hoes}.

There exists a good agreement between calculated $R$-matrix functions and the corresponding astrophysical factors for all cases presented in left panels (a), (b), and (c) as a result of fit procedure adjusting parameters (the reduced $\gamma$s widths) of each resonance, using as initialization parameters those reported in \citet{HEI08}. Although a similarly good agreement is reached between curves and THM indirect data in right panels for resonances above $500\;{\rm keV}$, the panel (e) best reproduces the threshold level, with a reduced $\tilde{\chi}^2=1.65$, using just one normalization parameter for energies between $-0.3$ and $1.2\;{\rm MeV}$. It is nevertheless important specify that this result does not represent the best fit \citep[the one found in][]{LAC13} for the low energy resonance, but we can note that the curve is well compatible, within uncertainties, with the THM $d^{2}\sigma/dE_{c.m.}d\Omega_{d}$ cross section. In fact, the idea is to determine absolute normalization using state-of-the-art resonance parameters and a single normalization constant over the whole energy region, as mandated by THM theory. The resonance located at $4.7\;{\rm keV}$ appears to be lower and larger than those the one shown in \citet{LAC13} modifying the contribution attributed to this state in the most relevant astrophysical region for stellar nucleosynthesis. In the other two cases, panels (d) and (f), the threshold resonance is always underestimated by calculations showing the need of a higher value for the ANC of the $1/2^{+}$ state. 
Therefore, we conclude that the combination of \citet{BAI73} and, at low energies, of \citet{DRO93} data represents the only direct data set compatible with the THM $S$-factor and the ANC of the threshold level measured in many experiments \citep{PEL08,GUO12,AVI15}. In the light of this concordance scenario, we will use this pool of direct measurements in the rest of the present work. Since in the energy region above $500\;{\rm keV}$ we find different parameters with respect to ones adopted in \citet{LAC13,LAC12} and \citet{HEI08}, variations in both cross section and astrophysical factor of the $^{13}{\rm C}(\alpha,n)^{16}{\rm O}$ reaction are expected, so we proceed to a reanalysis of the THM data to supply a new recommended reaction rate based on this internally consistent data set. A significant improvement in systematic errors is expected since the different approaches have different possible sources of systematic errors.

In Fig. \ref{hoes}, the angular integrated $d^2\sigma/dE_{c.m.}d\Omega_{d}$ cross section is displayed as full symbols, with the horizontal error bars defining the $^{13}{\rm C}-\alpha$ relative-energy binning. The vertical error bars of experimental data account for statistical, angular integration, and background subtraction uncertainties \citep[see ][for more details]{LAC13}. The presence of three peaks in the $^{13}{\rm C}-\alpha$ relative energy spectrum is clearly shown in Figure \ref{hoes} in the energy range between $\sim-300$ and $1200\;{\rm keV}$. The first resonance on the left corresponds to the $1/2^{+}$ excited states of $^{17}{\rm O}$ at $6.363\;{\rm MeV}$ \citep{FAE15}, again $4.7\;{\rm keV}$ over the $\alpha$-threshold and about $7.4\;{\rm keV}$  higher than the value suggested in \citet{TIL93} (the small discrepancy comes from a different value of the threshold in the two works), for long time considered to be the reference paper \citep[original evaluation by ][]{FOW73}. Moreover, arrows in Fig. \ref{hoes} suggest that, a priori, each of the two peaks above $500\;{\rm keV}$ refer to two separate, but with similar $E_r$, resonances: $7.165\;{\rm MeV}$, $7.248\;{\rm MeV}$, $7.380\;{\rm MeV}$ and $7.381\;{\rm MeV}$, respectively \citep[whose energies are taken from \citet{FAE15} and essentially confirmed by ][considering a deviation of a few ${\rm keV}$]{TIL93}. The central red line shown in Fig. \ref{hoes} was obtained assuming the same parameters used in panels (b) and (e) of Fig. \ref{dir-ind} for all resonances. In order to comprise also the possible interference between states of $^{17}{\rm O}$ with the same spin-parity $J^{\pi}$, we considered the interfering effect of $3/2^{+}$ sub-threshold resonance at $5.931\;{\rm MeV}$ \citep{FAE15}, together with the $7.248\;{\rm MeV}$ state. The upper and lower limit (delimiting the red region in Fig. \ref{hoes}) account for both statistical, connected to the scatter of data points below $500\;{\rm keV}$, and normalization uncertainties due to fitting procedure of higher energy data. Parameters used in calculations of this paper allow us to reproduce the observable partial width of the $4.7\;{\rm keV}$ resonance, $\Gamma_{n}^{1/2^{+}}=136\pm5\;{\rm keV}$ as suggested by \citet{FAE15}, and the ANC value of $3.6\pm0.7\;{\rm fm}^{-1}$ as in \citet{AVI15}.

\begin{figure}
\includegraphics[width=0.85\columnwidth]{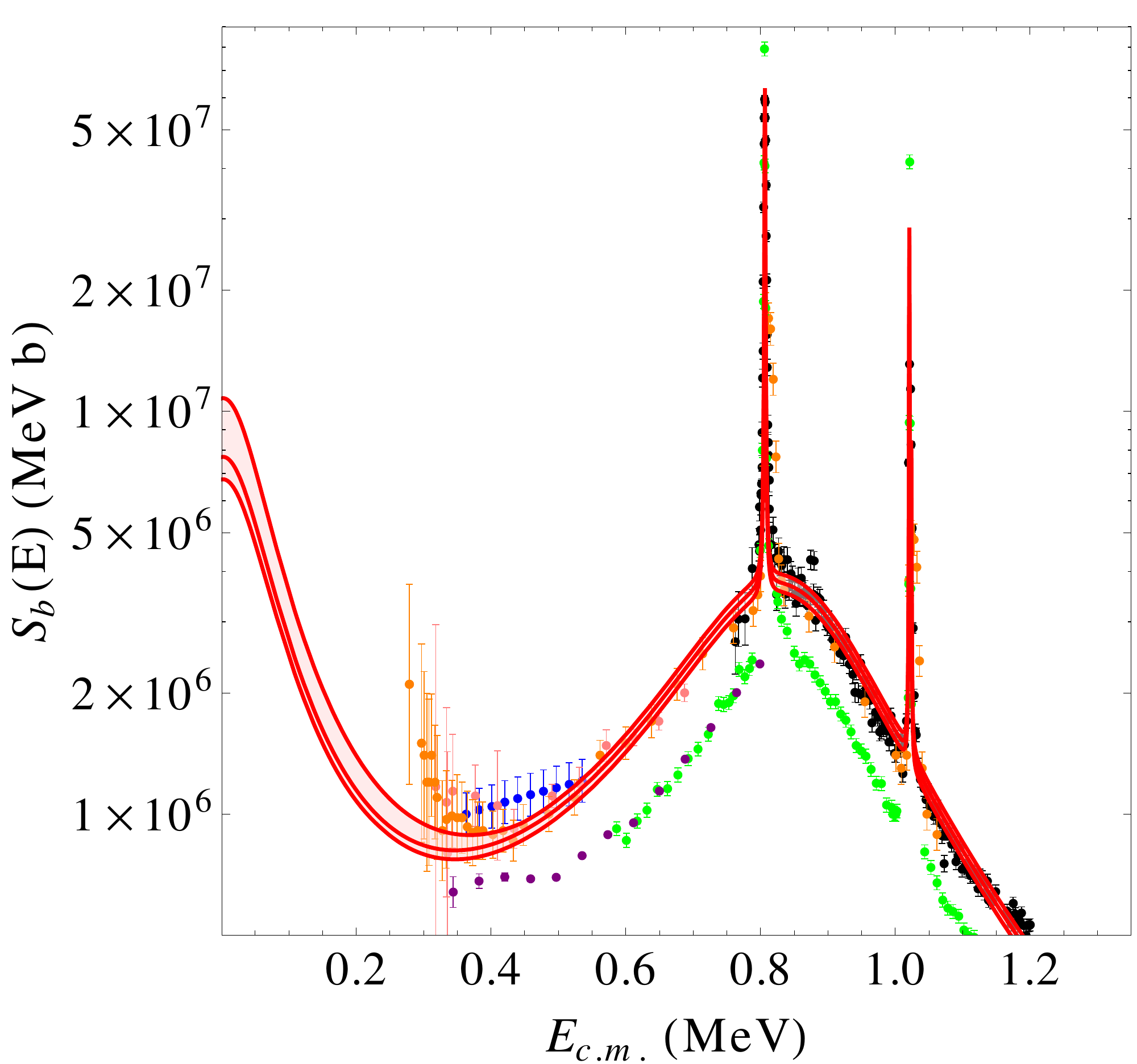}
\caption{$R$-matrix astrophysical factor (central red curve) calculated adopting resonance parameters used in panels (b) and (e) of Fig. \ref{dir-ind}. Both normalization and statistical uncertainties are delimited by the lower and upper red lines. At high energy uncertainties is assumed to be about $5\%$ in order to reproduced the average errorbar of \citet{BAI73} data. Direct data by \citet{DAV68,BAI73,KEL89,DRO93,HAR05,HEI08} are representing by blue, black, purple, orange, green, and pink solids points, respectively, without assuming any normalization.\label{astro-fac}}
\end{figure}

\section{Astrophysical Factor \label{sec-astro-fac}}
\begin{figure}
\includegraphics[width=0.85\columnwidth]{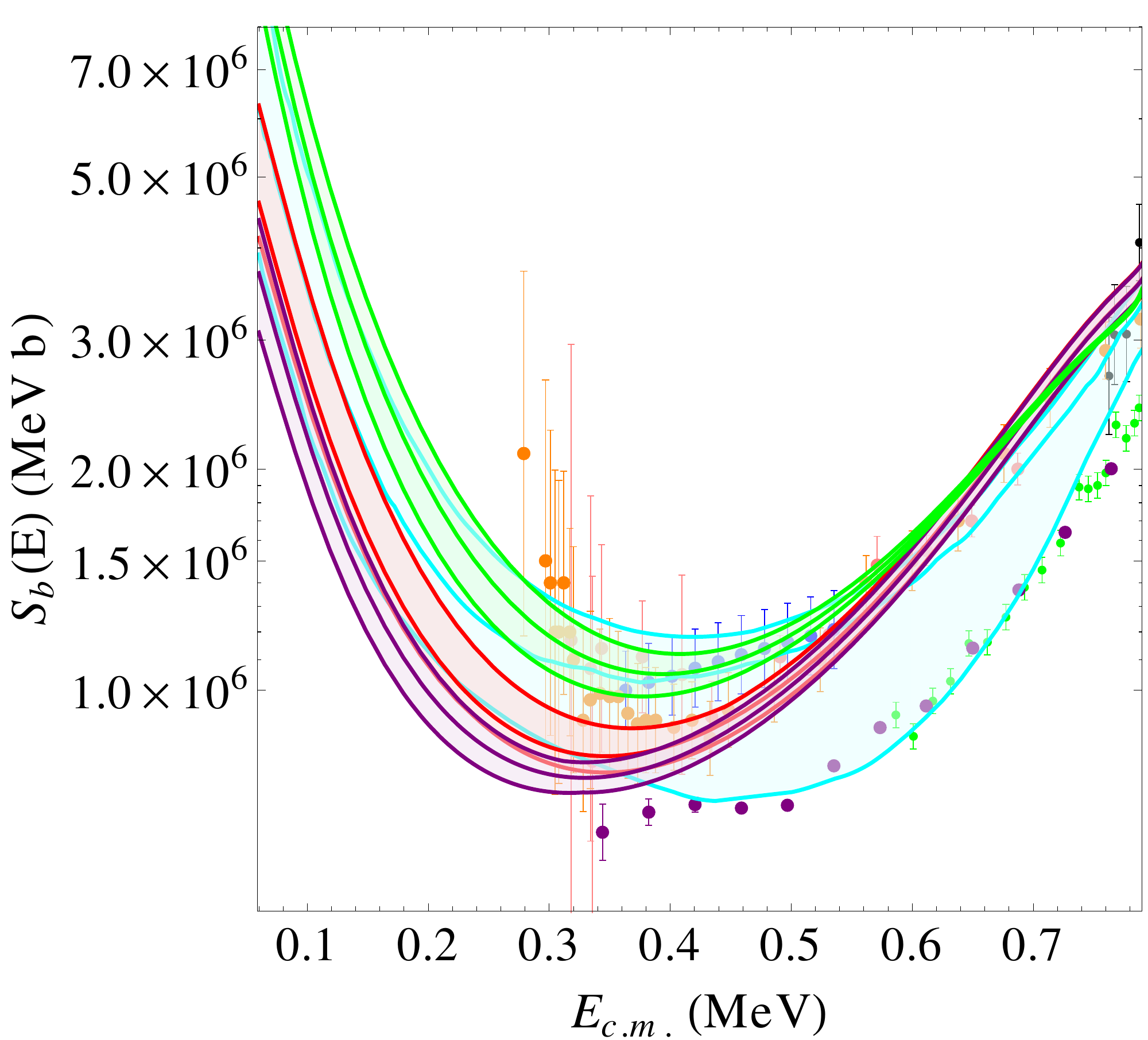}
\caption{Comparison between $S(E)$-factor calculated in this paper (red band) with recent indirect determinations by \citet{LAC13} and \citet{AVI15} (green and purple band, respectively). The cyan band, instead, shows the astrophysical factor and the corresponding uncertainties suggested by NACRE II compilation \citep{XU13}. For ease of comparison, the same data set of Fig. \ref{astro-fac} is shown in the low-energy region between $0.06$ and $0.8\;{\rm MeV}$ where the contribution of the $1/2^+$ state is more effective.\label{astro-fac-comparison}}
\end{figure}

The comparison of the THM $S(E)$-factor for the $^{13}{\rm C}(\alpha,n)^{16}{\rm O}$ reaction with the direct data from \citet{DAV68,BAI73,KEL89,DRO93,HAR05,HEI08} is shown Fig. \ref{astro-fac}. Direct measurements just mentioned, in contrast with \citet{LAC12,LAC13}, were not renormalized to the astrophysical factor of \citet{HEI08} to emphasize the differences among different datasets present in literature. For experimental measurement data we adopt the same symbols and colors that we used in Fig. \ref{dir-ind}. The astrophysical factor was obtained using a standard $R$-matrix code \citep{LAN58} adopting the same reduced widths appearing in the modified $R$-matrix approach. We account for all the resonances occurring inside the energy range between $-0.3$ and $1.2\;{\rm MeV}$, also including those states influencing the off-energy-shell $S(E)$-factor through their tails or interference with other resonances; in particular, we considered the interfering effect due to sub-threshold resonance at $5.931\;{\rm MeV}$ together with the $7.248\;{\rm MeV}$ state having the same spin-parity value $3/2^{+}$, as for HOES calculations. In Fig. \ref{astro-fac}, the result is shown by the central red line representing the best fit curve, while the red band, delimited by upper and lower red lines, account for statistical, normalization and data reduction uncertainty. As already discussed in section \ref{sec-norm}, a very good agreement exists between experimental data by \citet{BAI73} and the THM $S(E)$-factor. At the same time, red line is compatible with \citet{DRO93} at low energy, while no match is possible with measurements performed by \citet{KEL89} and \citet{HAR05}, showing a remarkably different absolute value. 

In order to provide a detailed comparison of the astrophysical factor obtained through the indirect TH approach with other determinations present in literature, Fig. \ref{astro-fac-comparison} displays only on the $60-790\;{\rm keV}$ low energy region, in the center-of-mass system, which is the most relevant for the astrophysical $s$-process scenario. In Fig. \ref{astro-fac-comparison} the THM $S(E)$-factor is compared only with the most recent evaluation of the $^{13}{\rm C}(\alpha,n)^{16}{\rm O}$ astrophysical factor. In particular, green and purple bands represents the $S$-factor by \citet{LAC13} namely, the precedent THM analysis, and the one extrapolated in \citet{AVI15}, respectively. There exist an agreement, inside error bars, between the $S(E)$-factor in \citet{AVI15} and the present one, but in \citet{LAC13} the band overestimates the new low-energy value, by a factor of $2.3$ for $E_{c.m.}<0.5\;{\rm MeV}$ because of the different ANC value assumed in the calculation. The $S(E)$-factor by NACRE II \citep{XU13} is shown by the cyan line, but there exist an agreement with the red curves only thank to large uncertainties (cyan band). In particular, we can also note that their lower limit is very close to \citet{HAR05} and \citet{KEL89} direct data. The highest uncertainties are, in fact, located in the normalization energy region, between $0.3$ and $0.7\;{\rm MeV}$, emphasizing once again the importance to discriminate among the absolute values of different direct datasets. In this paper, we approached and tried to solve this issue showed by NACRE II compilation considerably reducing the error associated with the normalization thanks to the concurrent application of ANC and THM.

\section{Electron Screening \label{sec-ele-scree}}
\begin{figure}
\includegraphics[width=0.85\columnwidth]{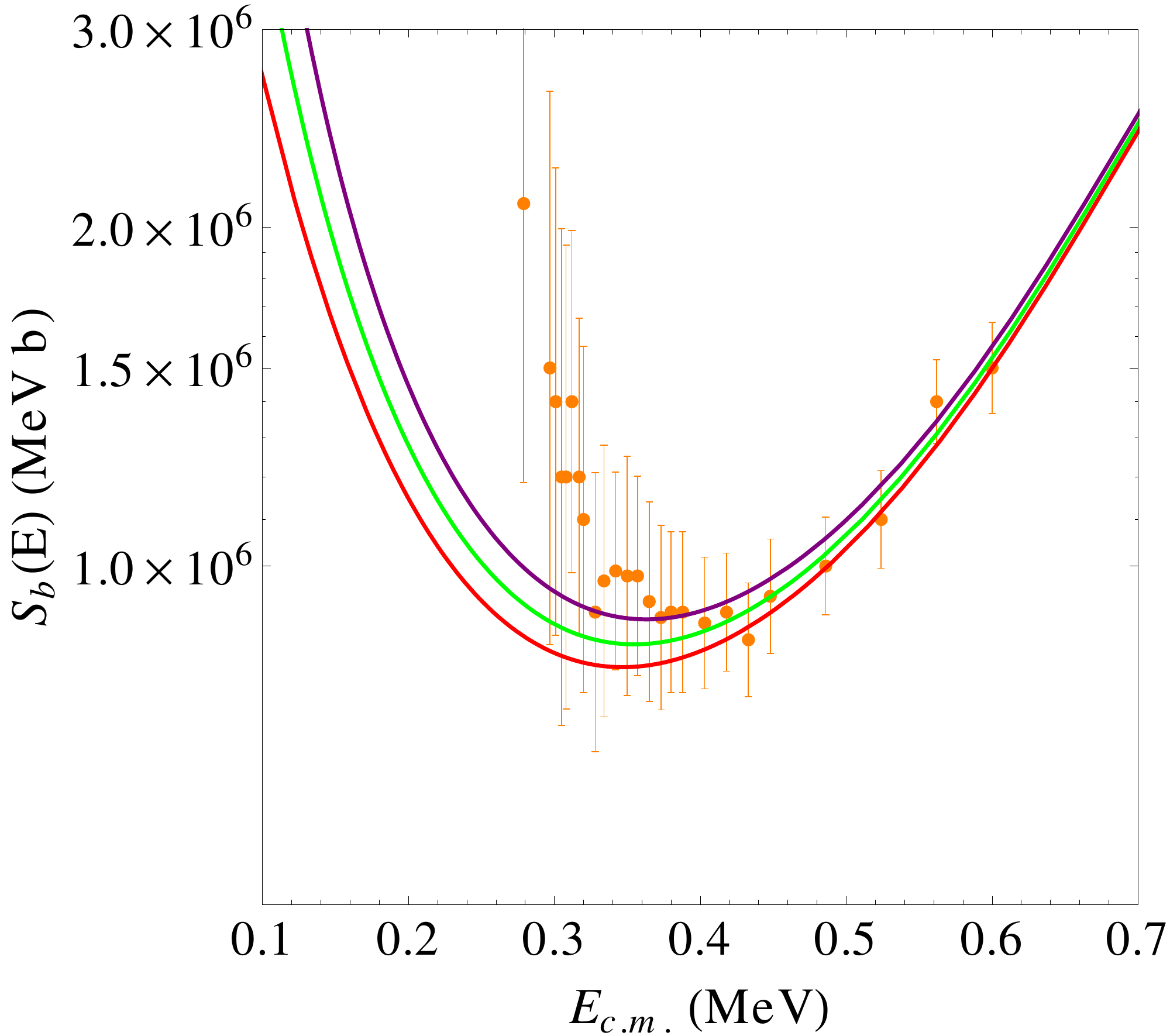}
\caption{Comparison between the bare-nucleus THM $S(E)$-factor (red line), just shown in Figure \ref{astro-fac}, and low-energy direct data measured by \citet{DRO93} (orange symbols). In particular, the \citet{DRO93} data set is taken from the \citet{ANG99} compilation, where no correction for the electron screening enhancement has been performed. A purple line is used for the screened $S$-factor, assuming the electron screening potential $U_e=2\;{\rm keV}$ suggested by \citet{ASS87}. The green curve is calculated using the adiabatic approach $U_e=0.937\;{\rm keV}$.\label{ele-scree}}
\end{figure}

\begin{figure}
\includegraphics[width=\linewidth]{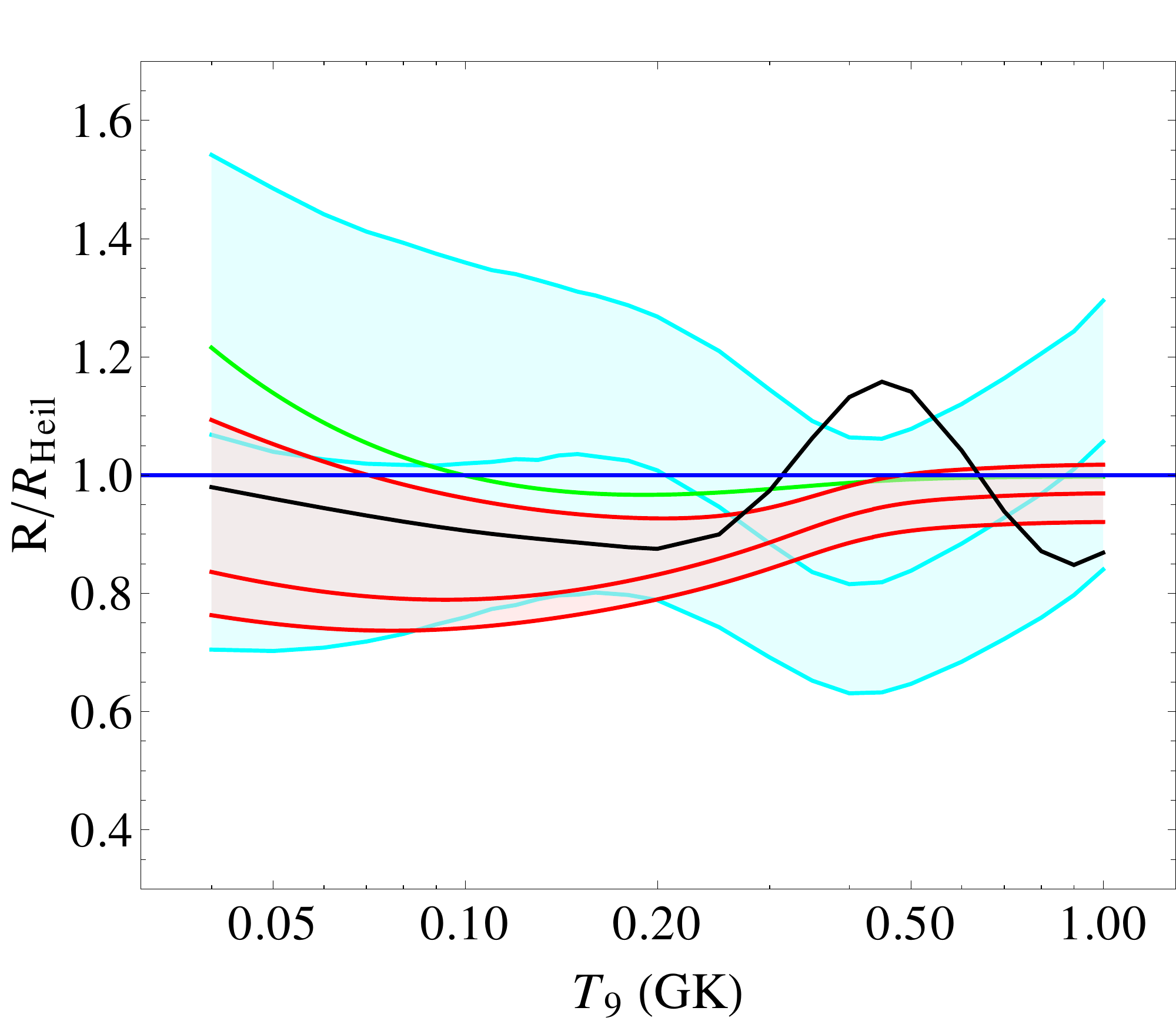}
\caption{Ratio between the reaction rate calculated using the THM astrophysical factor in Figure \ref{astro-fac} (red band) compared to the one of \citet{HEI08} (blue horizontal line), as a function of the temperature expressed in units of $10^9\;{\rm K}$. In the same reference, a green and a black curves represent the rates by \citet{LAC13} and \citet{DRO93}, respectively, while the cyan band shows the case of NACRE II \citep{XU13} and the corresponding uncertainties. (Color online.)\label{rate}}
\end{figure}

The determination of the $^{13}{\rm C}(\alpha,n)^{16}{\rm O}$ astrophysical $S(E)$-factor at very low energies by direct approaches strongly depends on the electron screening effect. Indeed, for nuclear cross section measured in the laboratory the projectile is usually in the form of an ion and the target is usually a neutral atom or a molecule surrounded by their electronic cloud. Following the basic idea \citep{ASS87,ROL88}, an impinging nucleus sees no repulsive nuclear Coulomb force until it penetrates beyond the atomic radius because of the electron cloud surrounding the target nuclei. Therefore, at low beam energies, the projectile is subject to a less repulsive potential because of the electron screening one $U_e$, basically the energy transfer from the atomic to nuclear degrees of freedom \citep{BRA90}. This results in an enhancement of the cross section relative to the value it would assume for fully ionized interacting particles. On the contrary, the THM measurement is not affected by the electron screening and it directly provides the bare-nucleus $S(E)$ factor (see red line in Fig. \ref{ele-scree}). It is not possible to attain a realistic estimate of the electron screening potential from the comparison of THM $S$-factor and direct data \citep[orange point ][without considering electron screening corrections]{DRO93} because of high experimental uncertainties affecting direct data. In this context, new high-accuracy direct measurements of the low-energy astrophysical factor of the ${}^{13}{\rm C}(\alpha,n){}^{16}{\rm O}$ can help us to better understand the electron screening effect for charged particle reactions. To estimate the electron screening effect, we have considered realistic electron screening potentials. In detail, green and blue lines represents the screened $S$-factor deduced adopting the adiabatic limit \citep[$U_e=0.937\;{\rm keV}$, ][]{BRA90} electron screening or assuming the large value ($U_e=2\;{\rm keV}$) suggested by \citet{ASS87}, respectively. Since data seem to suggest a $U_e$ value larger than the adiabatic limit, the case of $^{13}{\rm C}(\alpha,n)^{16}{\rm O}$ reaction might confirm what several measurements have demonstrated in the past, namely, that the experimental $U_e$ can be larger than the adiabatic limit \citep[see for instance ][ and references therein]{LAC05}.

\section{Reaction Rate \label{sec-rate}}

The thermonuclear rate for the $^{13}{\rm C}(\alpha,n)^{16}{\rm O}$ reaction was evaluated by means of standard equations \citep{ILI07} using the $S(E)$-factor measured through the THM devoid of electron screening effects (see Fig. \ref{astro-fac}). Table \ref{rate-tab} contains the adopted rate with the corresponding upper and lower limits in columns two, three, and four, respectively, as a function of the temperature expressed in ${\rm GK}$ (first column). The fifth column contains the exponents of the power-ten factor that is common to the three previous columns.

\begin{table*}
\begin{center}
\caption{Table of coefficients for the analytical approximation of the $^{13}{\rm C}(\alpha,n)^{16}{\rm O}$ reaction rate.\label{tab-coef}}
\begin{tabular}{ccccc}
\hline\hline
\multicolumn{2}{c}{\multirow{2}{*}{$a_{ij}$}} & & $j$ & \\
\cline{3-5}
 & & $1$ & $2$ & $3$ \\
\hline
\multicolumn{1}{c|}{} & $1$ & $+0.606751\times 10^{+2}$ & $-0.113003\times 10^{+2}$ & $+0.230810\times 10^{+2}$ \\ 
\multicolumn{1}{c|}{} & $2$ & $-0.487943\times 10^{-2}$ & $-0.109950\times 10^{+1}$ & $-0.675038\times 10^{+1}$ \\ 
\multicolumn{1}{c|}{} & $3$ & $-0.306167\times 10^{+2}$ & $-0.733349\times 10^{+0}$ & $-0.637515\times 10^{+1}$ \\ 
\multicolumn{1}{c|}{$i$} & $4$ & $-0.410653\times 10^{+2}$ & $+0.255634\times 10^{+2}$ & $-0.515203\times 10^{+1}$ \\ 
\multicolumn{1}{c|}{} & $5$ & $+0.238702\times 10^{+2}$ & $-0.129589\times 10^{+1}$ & $+0.169689\times 10^{+1}$ \\ 
\multicolumn{1}{c|}{} & $6$ & $-0.810502\times 10^{+1}$ & $-0.669953\times 10^{+1}$ & $-0.333333\times 10^{+1}$ \\ 
\multicolumn{1}{c|}{} & $7$ & $+0.383349\times 10^{+1}$ & $+0.910751\times 10^{+1}$ & $+0.207363\times 10^{+1}$ \\
\hline
\end{tabular}
\end{center}
\end{table*}

\begin{table*}
\begin{center}
\caption{Reaction rate of the $^{13}{\rm C}(\alpha,n)^{16}{\rm O}$ reaction. The recommended value, upper, and lower limits are displayed between $T_9 = 0.04$ and $1.00$, covering the astrophysical interesting region and neglecting values with power of $10$ less than $-30$.\label{rate-tab}}
\begin{tabular}{ccccc}
\hline\hline
 & & Reaction Rate & & \\ 
 & & $({\rm cm}^3\; {\rm mol}^{-1}\; {\rm s}^{-1})$ & & \\
\cline{2-4}
$T_9\;{\rm (GK)}$ & Adopted & Upper & Lower & Power of $10$\\
\hline
$0.04$ & $2.278$ & $2.978$ & $2.079$ & $-24$ \\
$0.05$ & $1.428$ & $1.843$ & $1.311$ & $-21$ \\
$0.06$ & $1.916$ & $2.441$ & $1.768$ & $-19$ \\
$0.07$ & $9.515$ & $11.98$ & $8.826$ & $-18$ \\
$0.08$ & $2.379$ & $2.960$ & $2.217$ & $-16$ \\
$0.09$ & $3.612$ & $4.444$ & $3.379$ & $-15$ \\
$0.10$ & $3.763$ & $4.580$ & $3.534$ & $-14$ \\
$0.11$ & $2.926$ & $3.523$ & $2.756$ & $-13$ \\
$0.12$ & $1.801$ & $2.147$ & $1.701$ & $-12$ \\
$0.13$ & $9.174$ & $10.83$ & $8.677$ & $-12$ \\
$0.14$ & $3.995$ & $4.671$ & $3.783$ & $-11$ \\
$0.15$ & $1.525$ & $1.767$ & $1.446$ & $-10$ \\
$0.16$ & $5.211$ & $5.985$ & $4.943$ & $-10$ \\
$0.18$ & $4.624$ & $5.227$ & $4.389$ & $-09$ \\
$0.20$ & $3.069$ & $3.420$ & $2.914$ & $-08$ \\
$0.25$ & $1.434$ & $1.554$ & $1.362$ & $-06$ \\
$0.30$ & $2.895$ & $3.087$ & $2.750$ & $-05$ \\
$0.35$ & $3.424$ & $3.621$ & $3.253$ & $-04$ \\
$0.40$ & $2.777$ & $2.925$ & $2.638$ & $-03$ \\
$0.45$ & $1.675$ & $1.761$ & $1.591$ & $-02$ \\
$0.50$ & $7.872$ & $8.271$ & $7.478$ & $-02$ \\
$0.60$ & $9.525$ & $10.00$ & $9.048$ & $-01$ \\
$0.70$ & $6.300$ & $6.616$ & $5.985$ & $+00$ \\
$0.80$ & $2.727$ & $2.864$ & $2.591$ & $+01$ \\
$0.90$ & $8.724$ & $9.161$ & $8.288$ & $+01$ \\
$1.00$ & $2.238$ & $2.350$ & $2.126$ & $+02$ \\
\hline
\end{tabular}
\end{center}
\end{table*}

The THM reaction rate can be described by the following analytical expression: $N_{A} \langle \sigma v \rangle = \sum\limits_{j=1}^3 {\rm exp} [a_{1j}+a_{2j}T_9^{-1}+a_{3j}T_9^{-1/3}+a_{4j}T_9^{1/3}+a_{5j}T_9+a_{6j}T_9^{5/3}+a_{7j}{\rm ln}(T_9)]$, where $T_9$ is the temperature in ${\rm GK}$ and $a_{ij}$ parameters of this expression, collected in Table \ref{tab-coef} for the recommended $^{13}{\rm C}(\alpha,n)^{16}{\rm O}$ reaction rate.

\begin{figure*}
\centering
\includegraphics[width=\linewidth]{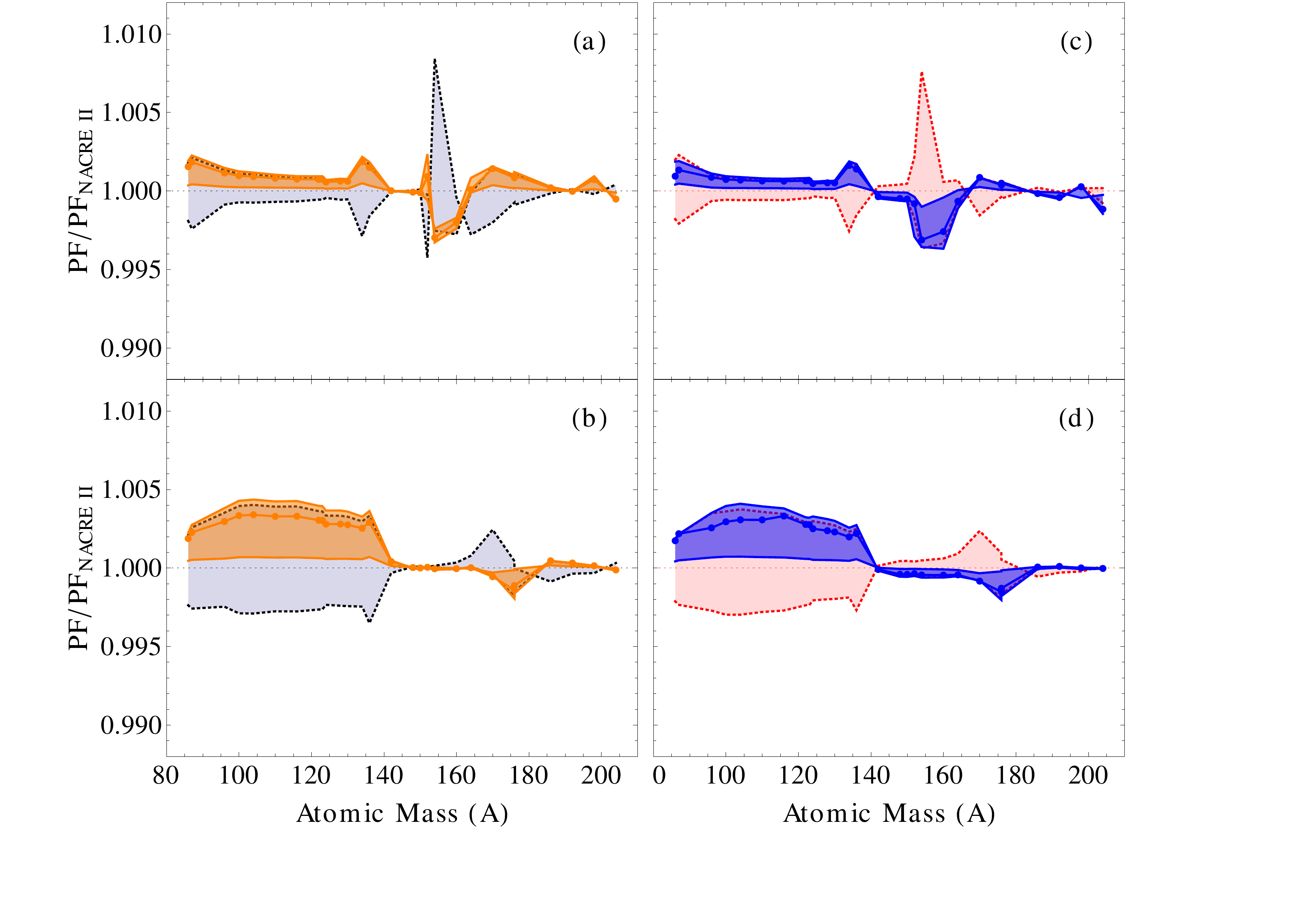}
\caption{\textit{Upper Panels}: Ratio between production factors (${\rm PF}$) of the main component ($A\geq86$) $s$-only nuclei calculated by NEWTON post-process code \citep{TRI14,TRI16} adopting the rate for the $^{13}{\rm C}(\alpha,n)^{16}{\rm O}$ reaction presented in this work (orange and blue bands) and the one suggested in NACRE II \citep[black and red shaded regions, ][]{XU13}. The horizontal dashed line corresponds to calculations performed adopting the recommended values of \citet{XU13} and it is considered to be the reference value (equal to $1$ in this representation).
The orange and black bands, in the \textit{left panel} (a), refer to a $^{13}{\rm C}$ reservoir of $5.0\times10^{-3}\;{\rm M}_{\odot}$ as used in \citet{TRI16}. In the \textit{right panel} (c), a different $^{13}{\rm C}$-pocket of about $3.5\times10^{-3}\;{\rm M}_{\odot}$ was used to calculate both red and blue bands. All curves in upper panels were calculated by means of stellar model of $1.5\;{\rm M}_{\odot}$ and almost solar metallicity ${\rm[Fe/H]}=-0.15$ experiencing nine thermal instabilities. \textit{Lower Panels}: As in the upper panels, but in the case of a star characterized by $3.0\;{\rm M}_{\odot}$ and about one third of solar metallicity ${\rm[Fe/H]}=-0.50$ experiencing eleven thermal instabilities. (Color online.) \label{astro}}
\end{figure*}

Fig. \ref{rate} show the reaction rate calculated from the THM $S(E)$-factor (red central line) and compared to the reference one \citep{HEI08} in the temperature range between $0.005$ and $1.0\;{\rm GK}$. The two rates are compatible almost everywhere, but the greater divergence is just located in the most interesting region for astrophysics, i.e. the discrepancy at $T_9= 0.09$ is around $12\%$. The red band represents the corresponding statistical and normalization uncertainties of the order of $+40\%$ and $-13\%$ at lower temperature. Higher differences are expected with the reaction rate suggested in \citet{LAC13} (green line), coming from the precedent analysis of THM data, because it is a factor $2.5$ larger than in \citet{HEI08} and the present work calculation at $5.0 \times 10^{7}\;{\rm K}$. The rates for $^{13}{\rm C}(\alpha,n)^{16}{\rm O}$ reaction by \citet{XU13} and \citet{DRO93} are also represented \citep[normalized to ][adopted value]{HEI08} in Fig. \ref{rate} with a cyan shaded region and a black curve, respectively. The rate from the present work agrees, within errors, with the updated NACRE II compilation \citep{XU13} recommended value in the temperature range of main astrophysical interest ($0.05 < T_9 < 0.30$). The great difference consists in the level of uncertainties, the upper limits of \citet{XU13} is $36\%$ higher than the corresponding recommended value, while in our case it is only $17\%$ at about $T_9=0.09$; on the other side, lower limits are $28\%$ and $7\%$, respectively, then the present rate is more accurate of about a factor of $4$. In other temperature regions, the NACRE II rates (both adopted and upper) are at most about $1.3$ times higher than the rate reported here, while the lower limits are even closer to each other. On the other side, the reaction rate by \citet{DRO93} significantly differs from the new THM one for temperatures higher than $0.3\;{\rm GK}$. These differences should be evaluated in all astrophysical sites mentioned in the first section for which the activation of the $^{13}{\rm C}(\alpha,n)^{16}{\rm O}$ reaction is relevant.

\section{Astrophysical Implications of New Reaction Rate \label{astro-cons}}
The main aim of this paper is to provide a rate for the $^{13}{\rm C}(\alpha,n)^{16}{\rm O}$ reaction decreasing systematic errors connected to the normalization between measurements performed through indirect techniques and those direct data-sets showing different absolute values. This, in fact, still represents the main  source of errors also in the NACRE II compilation \citep{XU13}. In order to verify the impact of the new rate for the $^{13}{\rm C}(\alpha,n)^{16}{\rm O}$ reaction on stellar nucleosynthesis, we performed some theoretical prediction for the specific case of the $s$-processing in AGB-LMSs. 

In this astrophysical scenario, the $^{13}{\rm C}(\alpha,n)^{16}{\rm O}$ is strictly connected to the amount of $^{13}{\rm C}$ locally produced in the ${\rm He}$-rich region of an AGB-LMS, as a consequence of some proton penetration during the phases subsequent to the development of thermal instabilities. The abundant $^{12}{\rm C}$ can interact with these protons producing fresh $^{13}{\rm C}$ through the reaction chain $^{12}{\rm C}(p,\gamma)^{13}{\rm N}(\beta^+\,\nu)^{13}{\rm C}$. Too efficient proton captures, indeed, can activate a full CN cycling, leading to the production of $^{14}{\rm N}$ through the reaction $^{13}{\rm C}(p,\gamma)^{14}{\rm N}$. This fact is of crucial importance because $^{14}{\rm N}$ is a very efficient absorber for neutrons, the main $n$-poison in typical AGB star conditions, which would inhibit the captures on heavier nuclei. At this point the star presents a $^{13}{\rm C}$-pocket embedded in a ${\rm He}$-rich environment, so that, when the temperature approaches $(0.9-1.0)\times 10^8\;{\rm K}$, the $^{13}{\rm C}(\alpha,n)^{16}{\rm O}$ reaction is activated, releasing neutrons. 

In most evolutionary codes this penetration is treated as a free parameter suggesting a depth down to $(0.5 - 1)\times10^{-3}\;{\rm M}_{\odot}$ \citep{GAL88,BIS10}. Other mechanisms were recently suggested to physically model the proton penetration in the ${\rm He}$-rich region and to consequently form a $^{13}{\rm C}$ reservoir. \citet{BAT16} suggested gravity waves and Kelvin-Helmoltz instabilities as possible cause of proton penetration, while hydrodynamical effects induced by convective overshooting at the border of convective envelope were considered by \citet{CRI11}. In \citet{PIE13} also the effects due to rotation in stars are considered showing that the Goldreich-Schubert-Fricke instabilities affect the formation of the $^{13}{\rm C}$ pocket. The possibility that magnetic buoyancy represents a physical mechanism suitable to produce an extended $^{13}{\rm C}$-pocket ($3.5-5 \times 10^{-3}\;{\rm M}_{\odot}$) with a rather flat $^{13}{\rm C}$ distribution was investigated in \citet{BUS07,TRI14,TRI16}. In this scenario, the H-burning re-ignition induces a low concentration of $^{13}{\rm C}$ (but with a profile extending to several $10^{-3}\;{\rm M}_{\odot}$), yielding a negligible abundance of the neutron-poison $^{14}{\rm N}$. The idea of a large hydrogen penetration was originally suggested to reproduce the spectroscopic observations in young open clusters by \citet{MAI12}, performed with new analysis methods and showing enhancements of $s$-element abundances with respect to the Sun. To discriminate between different scenarios for the $^{13}{\rm C}$-pocket formation or argue in favour of one reservoir profile with respect to another is beyond the aim of this paper. So, we only adopted the mechanism for proton penetration suggested in \citet{TRI16} and performed $s$-process nucleosynthesis calculations by means of NEWTON post-process code \citep{BUS99,TRI14,TRI16}, containing a detailed
network of more than 400 isotopes (from ${\rm He}$ to ${\rm Bi}$) connected by $\alpha$-, $p$- and $n$- reactions and weak interactions, in order to assess and understand the impact of the new rate for the $^{13}{\rm C}(\alpha,n)^{16}{\rm O}$ reaction. 

Fig. \ref{astro} shows the ratios between production factors with respect to solar abundances of neutron rich nuclei belonging to the $s$-process main component, that is considered to start at $A=86$ and corresponds to nuclides of elements between strontium and bismuth in LMSs, adopting the reaction rate presented in this paper and the one of NACRE II \citep{XU13} assumed as the reference one, i.e. black and red dashed horizontal lines means no variations between two scenarios. We can note from Fig. \ref{rate} that in the interesting energy region for AGB-LMSs ($0.8-1.0\times10^8\;{\rm K}$) the $^{13}{\rm C}(\alpha,n)^{16}{\rm O}$ reaction rates by \citet{HEI08}, \citet{LAC13}, and \cite{XU13} are very similar, so $s$-nucleosynthesis results are expected to be almost comparable. Only isotopes that are produced exclusively by slow neutron captures (the so-called $s$-only nuclei), being shielded against the fast decays of the $r$-process by stable isotopes, are reported in Fig. \ref{astro}. 

Calculations in the upper panels (a and c) of Fig. \ref{astro} refer to a low-mass star of $1.5\;{\rm M}_{\odot}$ and almost solar metallicity \citep[${\rm [Fe/H]}=-0.15$, assuming as reference ][]{LOD09} experiencing nine thermal instabilities. We considered this stellar model because in \citet{TRI14} it was noticed that the results of the averaging procedure on LMS models in the range between $1.5$ and $3.0\;{\rm M}_{\odot}$ are very similar to the calculations of the individual model at smaller mass. This fact was the consequence of the Salpeter's initial mass function which favours lower masses in the weighting operation. For these reasons, this single model represents a reasonable example in order to reproduce the flat behaviour of solar distribution for $s$-process nuclei \citep{TRI14,TRI16}. In Fig. \ref{astro} we used bands in order to take into account also upper and lower limits of reaction rates of this paper and by \citet{XU13}. The production factors of the selected isotopes in the ${\rm He}$-rich region of the star show that variations are very limited and they are always lower than $1.0\%$, so the effect of reaction rate of Tab. \ref{rate-tab} is essentially negligible with respect to calculations obtained with \citet{XU13} reaction rate. The only peculiar variations are located in the region of gadolinium isotopes of panel (a): there exists a production of $^{152}{\rm Gd}$ and a destruction of $^{154}{\rm Gd}$, both around $5-8$ parts per thousands, with the new rate with respect to the \citet{XU13} one. However, the improvement in the accuracy of the predictions, as demonstrated by the shrinking of the uncertainty bands, is apparent, bespeaking the power of the approach outlined in this work.

Orange and blue curves correspond to the deeper ($5.0\times10^{-3}\;{\rm M}_{\odot}$) and more limited ($3.5\times10^{-3}\;{\rm M}_{\odot}$) proton penetration, and consequently for the extension of the $^{13}{\rm C}$ reservoirs suggested in \citet{TRI16}, respectively. The blue line, corresponding to a pocket with a smaller amount of $^{13}{\rm C}$, shows lower production of $s$-only nuclei, but variations are still of the order of few per thousands. Differences between blue and orange curves are so limited that, not also being the aim of our article as we have already mentioned before, we do not conclude in favour of one $^{13}{\rm C}$ reservoir with respect to the other, taking into account present-day accuracy of observational data. We underscore that the PFs in Fig. \ref{astro} are likely to be more sensible to the uncertainties affecting the reaction rate with respect to the total extension, in mass, of the two $^{13}{\rm C}$ pockets described above. Nonetheless, the mechanism for proton injection into the ${\rm He}$-rich shell still represents a debated open issue strongly related to start-time, rate, profile, and mass of the $p$-penetration. The formation of $^{13}{\rm C}$ reservoirs is then model-dependent because only full 3-dimensional magneto-hydrodynamical simulations could really help making progress in the understanding of the pocket and different choices for the quantities mentioned above could result in higher variations for the production of $s$-nuclei.

The lower panels (b and d) of Fig. \ref{astro} show the same calculations of upper ones (a and c, respectively), but adopting a different stellar model of $3.0\;{\rm M}_{\odot}$ and one third of solar metallicity (${\rm [Fe/H]}=-0.50$) experiencing eleven thermal instabilities. We performed calculations for different stellar models to separate the effect of the rate by the one connected to the variation of stellar parameters. In the $3.0\;{\rm M}_{\odot}$ case, the neutrons from the $^{22}{\rm Ne}$ source are important and produced at a rather high neutron density, while in the lower-mass models the $^{13}{\rm C}$ source always dominates. For the sake of simplicity, we use the same colours and symbols as before. As in previous cases, the $3.5\times10^{-3}\;{\rm M}_{\odot}$ $^{13}{\rm C}$ reservoir (blue curve) shows a more limited production for the $s$-only nuclei with comparison to the more extended one ($5\times10^{-3}\;{\rm M}_{\odot}$, orange line). Comparing upper (a and c) and lower (b and d) panels of Fig. \ref{astro}, the $3.0\;{\rm M}_{\odot}$ and $1/3$ of solar metallicity star expects higher variations for $s$-only nuclei up to barium (heavy $s$ peak) with respect to star of $1.5\;{\rm M}_{\odot}$ and almost solar metallicity.

On the contrary, comparing left (a and b) and right (c and d) panels on the same row we can note the $^{13}{\rm C}$-pocket weakly influences the production and variations of production factor for $s$-only nuclei are smaller than those expected by changing the reference rate in the nucleosynthesis code. In particular, black and red curves, both obtained selecting the \citep{XU13} reaction rate, are very similar to each other, as well as in the case of orange and blue lines, representing the calculation performed with the $^{13}{\rm C}(\alpha,n)^{16}{\rm O}$ rate presented in this paper. In the cases discussed above, the production of the $s$-isotopes seems to be slightly more dependent on reaction rate choice with respect to the extension of the $^{13}{\rm C}$, stressing on the importance of well known nuclear inputs, decreasing normalization uncertainties and determining a way to define the absolute value of measured astrophysical factor, in stellar nucleosynthesis. 

Moreover, the new calculation presented in this paper shows a reduction of the rate for the $^{13}{\rm C}(\alpha,n)^{16}{\rm O}$ reaction with respect to our \citep{LAC12,LAC13} or \citet{HEI08} determinations in the most interesting region for astrophysics. These conditions could leave a bigger amount of unburned $^{13}{\rm C}$ \citep{CRI11,GUO12} than the previous predictions. Some $^{13}{\rm C}$, survived from the inter-pulses stage, is engulfed in the convective shell and burns at higher temperatures, about $1.5\times 10^{8}\;{\rm K}$, providing another neutron burst at higher densities via the $^{13}{\rm C}(\alpha,n)^{16}{\rm O}$ reaction itself. The percentage of $^{13}{\rm C}$ depends on the metallicity and initial mass of the star. As a consequence, the activation of some $s$-process branchings could modify production and/or destruction of branching-dependent isotopes. Verification of the results presented here using independent nucleosynthesis codes is highly desirable.

\section{Conclusions \label{sec-conc}}
In this paper, we present a new approach that, overturning the usually normalization procedure, strongly constraints absolute value of $S(E)$-factor for direct measurements starting from indirect experimental data; thus turning one of the most crucial drawbacks of indirect techniques into an advantage. This procedure, a priori, can be used for all those nuclear reactions of astrophysical interest which are characterized by great uncertainties at low energy and/or whose direct measurements show different absolute values for the corresponding astrophysical factor, for which ANCs of one of more resonances are measured with high accuracy.

We focused our attention on the case of the $^{13}{\rm C}(\alpha,n)^{16}{\rm O}$ reaction because it represents the main neutron source for AGB-LMSs, yet, it is affected by large systematic errors due to the spread in absolute normalization even at high energies, as also recently confirmed by the NACRE II compilation \citep{XU13}. In particular, by implementing the recent and precise ANC ($3.6\pm0.7\;{\rm fm}^{-1}$) calculated by \citet{AVI15} and the full width ($136\pm5\;{\rm keV}$) of \citet{FAE15} for the threshold resonance that strongly affect the $^{13}{\rm C}(\alpha,n)^{16}{\rm O}$ astrophysical factor in the corresponding Gamow window into a modified R-matrix fit of THM data, it was possible to define an absolute normalization of direct data by \citet{BAI73} are the only ones defining a concordance scenario together with ANC measurements and THM data. On the contrary, other datasets, \citet{HAR05} plus \citet{KEL89} or the combination of \citet{DRO93} with \citet{DAV68} and \citet{HEI08}, seem to suggest higher values for the ANC, as just observed in \citet{LAC12,LAC13}, not compatible with presently accepted values. 

The fact that \citet{FAE15} also predicted a small shift of about $7\;{\rm keV}$ towards positive $E_{c.m.}$ for the center of the resonance near the $\alpha$-threshold did not result in substantial changes, given energy resolution characterizing the THM cross section \citep{LAC13}. However, as an important consequence, the procedure to calculate the ANC of the same resonance must be changed in comparison to what was previously done in \citet{LAC12,LAC13}. As just previously discussed, the calculated value for the squared Coulomb-modified ANC $\left( \tilde{C}_{\alpha{}^{13}{\rm C}}^{{}^{17}{\rm O}(1/2^+)}\right)^2$ is $3.6\pm0.7\;{\rm fm}^{-1}$ and then well agrees with the determinations of \citet{AVI15} and other works \citep{PEL08,GUO12}.

In the light of this concordance scenario, the THM cross section was fitted to determine the resonance parameters, used to determine the astrophysical factor for the $^{13}{\rm C}(\alpha,n)^{16}{\rm O}$ reaction in the energy range between $0$ and $1.2\;{\rm MeV}$. The resulting $S(E)$-factor is lower than in \citet{LAC13} and substantially agrees, inside errorbars, with ones shown in \citet{AVI15} and in \citet{XU13}. We then calculated the recommended reaction rate by means of standard equations providing both a tabular list and an analytical formula. Adopting the new rate suggested in the present paper as input for the NEWTON code \citep{TRI14,TRI16} for the $s$-process nucleosynthesis in LMSs we expect only limited variations, less than few per thousand, for those nuclei whose production is considered to be totally due to slow neutron captures.

\begin{acknowledgments}
We thank the Referee for helping us improving the manuscript. We are grateful to M. Busso and S. Palmerini for collaboration and very useful discussion on $s$-process nucleosynthesis and A. M. Mukhamedzhanov and B. Irgaziev for what concerns ANC calculation. O.T. thanks both Department of Physics and Geology at the University of Perugia and Group 3 of INFN, section of Perugia, for financial support.
\end{acknowledgments}

\appendix
\section{EVALUATION OF THE ANC AT POSITIVE ENERGIES \label{sec-ANC}}
In \citet{LAC12,LAC13}, we limited our discussion to the extraction of the ANC for the $^{13}{\rm C}(\alpha,n)^{16}{\rm O}$ reaction from THM data in the specific case of a bound state (or ${\rm bs}$). As shown in \citet{LAC13} in the case of a virtual or real decay $B \rightarrow a+A$, in the channel with relative orbital angular momentum $l_B$, total angular momentum $j_B$ of $a$, total angular momentum $J_B$ of the system $a + A$, and when dealing with sub-threshold resonances, the squared ANC ($( C^{B}_{aA\,l_{B}\,j_{B}\,J_{B}})^2$, or $C^2$ for short) can be derived from the reduced width for the bound state $B=(aA)$ as follows \citep{MUK12}:
\begin{equation}
\centering
C^{2}=\left(\frac{\hbar^{2} W(R_{aA})^{2}}{2\mu\,R_{aA}\,\gamma^2}+\int_{R_{aA}}^{\infty}\vert W(r)^2\vert dr\right)^{-1}\;,
\label{anc-1}
\end{equation}
where $W(R_{aA})=W^2_{-\eta^{\rm bs}_{aA}\,l_{B}+1/2}$ is the Whittaker function and $R_{aA}$ the channel radius. Moreover, $\mu_{aA}$ and $\eta^{\rm bs}_{aA}$ are the reduced mass of the $a+A$ system and the Coulomb parameter for the bound state, respectively. The equation above has to be modified as for bound states at negative energies penetrability is zero, but the shift function can still be defined as the logarithmic derivative of the Whittaker function \citep{THO09}.


In the case of $^{13}{\rm C}(\alpha,n)^{16}{\rm O}$ reaction, the $1/2^{+}$ resonance state is close to the threshold and the Coulomb factor can not be neglected defining the modified ANC ($\tilde{C}$). There is one very important reason why we can not exclude the Coulomb factor; the renormalization does not change the reduced width, but allows one to operate with more reasonable ANCs values than the standard ones. At very low energy, in collision of charged particles, the Coulomb factor is so huge that ANC below threshold has very large value $\Gamma(l_b+1+\eta^{\rm bs}_{aA})$ (bound state).

However, for capture to an unbound state, we have to use the following formula \citep[Eq. 7 in ][]{MUK12}: 
\begin{equation}
\centering
C^{2}=(-1)^{l_{B}}e^{\pi \eta_{aA}} e^{2 i \delta^{p}_{l_B\,j_B\,J_B}\left(k_{aA(R)} \right) } \frac{\mu_{aA} \Gamma_{aA}}{k_{aA(R)}}\;,
\label{anc-2}
\end{equation}
where $\delta^{p}_{l_B\,j_B\,J_B}\left(k_{aA(R)}\right) $ is the potential (non-resonance) scattering phase shift at the real resonance relative momentum $k_{aA(R)}$, $\Gamma_{aA}$ represents the partial width of the excited state of $^{17}{\rm O}$ mentioned above, and $\eta_{aA}$ is the Sommerfeld parameter. Equation \ref{anc-2} must be corrected by the factor $\Gamma(l_{b}+1+i\eta_{aA})$, since the ANC for low energy collisions of charged particle is exceedingly small.


\providecommand{\noopsort}[1]{}\providecommand{\singleletter}[1]{#1}%

\end{document}